\documentclass[journal=nalefd,manuscript=article,layout=traditional]{achemso}

\usepackage[version=3]{mhchem} 
\usepackage{natbib}
\usepackage{color}

\author{Yen-Hui Lin}
\affiliation{Department of Physics, National Tsing Hua University, 30013 Hsinchu, Taiwan}
\author{Chia-Hsiu Hsu}
\affiliation{Department of Physics, National Sun Yat-sen University, Kaohsiung 804, Taiwan}
\author{Iksu Jang}
\affiliation{Department of Physics, National Tsing Hua University, 30013 Hsinchu, Taiwan}
\author{Chia-Ju Chen}
\affiliation{Department of Physics, National Tsing Hua University, 30013 Hsinchu, Taiwan}
\author{Pok-Man Chiu}
\affiliation{Department of Physics, National Tsing Hua University, 30013 Hsinchu, Taiwan}
\author{Deng-Sung Lin}
\affiliation{Department of Physics, National Tsing Hua University, 30013 Hsinchu, Taiwan}
\alsoaffiliation{Center for Quantum Technology, National Tsing Hua University, Hsinchu 30013, Taiwan}
\author{Chien-Te Wu}
\affiliation{Department of Electrophysics, National Chiao Tung University, Hsinchu, 30010, Taiwan}
\author{Feng-Chuan Chuang}
\email{fchuang@mail.nsysu.edu.tw}
\affiliation{Department of Physics, National Sun Yat-sen University, Kaohsiung 804, Taiwan}
\alsoaffiliation{Physics Division, National Center for Theoretical Sciences, Hsinchu 30013, Taiwan}
\author{Po-Yao Chang}
\email{pychang@phys.nthu.edu.tw}
\affiliation{Department of Physics, National Tsing Hua University, 30013 Hsinchu, Taiwan}
\author{Pin-Jui Hsu}
\email{pinjuihsu@phys.nthu.edu.tw}
\affiliation{Department of Physics, National Tsing Hua University, 30013 Hsinchu, Taiwan}
\alsoaffiliation{Center for Quantum Technology, National Tsing Hua University, Hsinchu 30013, Taiwan}

\title{
Proximity-Effect-Induced Anisotropic Superconductivity in Monolayer Ni-Pb Binary Alloy}

\abbreviations{IR,NMR,UV}

\begin{document}


\begin{abstract}

Proximity effect facilitates the penetration of Cooper pairs that permits superconductivity in normal metal, offering a promising approach to turn heterogeneous materials into superconducting and develop exceptional quantum phenomena. Here, we have systematically investigated proximity-induced anisotropic superconductivity in monolayer Ni-Pb binary alloy by combining scanning tunneling microscopy/ spectroscopy (STM/STS) with theoretical calculations. By means of high temperature growth, the $(3\sqrt{3}\times3\sqrt{3})R30^{o}$ Ni-Pb surface alloy has been fabricated on the Pb(111), where the appearance of domain boundary as well as lattice transformation are further corroborated by the STM simulations. Given the high spatial and energy resolution, tunnelling conductance  ($\mathrm{d}I/\mathrm{d}U$) spectra have resolved a reduced but anisotropic superconducting gap $\Delta_{NiPb} \approx 1.0$ meV, in stark contrast to the isotropic $\Delta_{Pb} \approx 1.3$ meV on the conventional Pb(111). In addition, the higher density of states at Fermi energy ($D(E_{F})$) of Ni-Pb surface alloy results in an enhancement of coherence peak height. According to the same $T_{c} \approx 7.1$ K with Pb(111) from the temperature dependent $\Delta_{NiPb}$ and a short decay length $L_{d} \approx $ 3.55 nm from the spatially monotonic decrease of $\Delta_{NiPb}$, both results are supportive for the proximity-induced superconductivity. Despite a lack of bulk counterpart, the atomic-thick Ni-Pb bimetallic compound opens a new pathway to engineer superconducting properties down to the low-dimensional limit, giving rise to the emergence of anisotropic superconductivity via proximity effect.

\end{abstract}
\section{Introduction}

By exploiting proximity effect\cite{PGdeGennes,JJHauser,CWJBeenakker,CLambert}, a normal metal in direct contact with a conventional \textit{s}-wave superconductor can become superconducting, which provides a novel method to artificially manufacture superconductivity in a wide variety of low-dimensional materials. For instance, the odd-frequency spin-triplet superconductivity has been realized in the thin layered ferromagnet/superconductor systems where the superconducting correlations remain coherent over a finite distance that enables the coexistence of ferromagnetism and superconductivity\cite{AIBuzdin,FSBergeret}. Besides ferromagnet/superconductor heterostructures, the two-dimensional (2D) spinless $\textit{p}_{x}+\textit{ip}_{y}$ superconductor can even be resembled by introducing Cooper pairs to chiral surface state of a 3D topological insulator (TI)\cite{LFu,XLQi}, which has been experimentally achieved in several composite systems involving TI thin films proximitized with superconductors in recent years\cite{JRWilliams,MVeldhorst,MXWang}. 

Accompanied by the great advancements in making nanometer scale materials superconducting via proximity effect, not only exceptional superconductivity, but also unique electronic properties have been developed in the reduced dimensions. Taking the aforementioned ferromagnet/superconductor systems for example, close to the ferromagnet/superconductor interfaces\cite{FSBergeret1,AKadigrobov,MEschrig}, singlet Cooper pairs can turn into spin-polarized to generate triplet supercurrents that leads to the minimized dissipation of Joule heating and the long spin lifetimes, having potential use for the spin transport applications\cite{RSKeizer,HYang,MEschrig1}. Likewise, the proximity-induced 2D spinless $\textit{p}_{x}+\textit{ip}_{y}$ superconductor exhibits the topological superconductivity, where the \textit{p}-wave-like pairing has been stabilized to host a new type of quasiparticle excitation, i.e., Majorana fermions (MFs), bound to the center of Abrikosov vortex cores\cite{JPXu,DFWang,TMachida}. Such Majorana bound states obey non-Abelian statistics in braiding processes of swapping vortices adiabatically, serving as an important candidate for topological qubits in realizing fault-tolerant quantum computation\cite{JAlicea,CWJBeenakker1,DAasen}. Besides the vortex cores, proximity-induced superconductivity along with strong spin-orbital interaction further supports MFs at the edges of semiconductor nanowire\cite{JDSau,VMourik,MTDeng}, magnetic atomic chain\cite{SNadjPerge,MRuby,HKim} and magnetic nanoisland\cite{APMorales,GCMenard} as well.

In this work, we have systematically investigated the proximity-induced superconductivity in monolayer Ni-Pb binary alloy by employing STM/STS together with theoretical calculations. By depositing Ni onto Pb(111) at elevated temperature, the $(3\sqrt{3}\times3\sqrt{3})R30^{o}$ Ni-Pb surface alloy has been fabricated with uniform Pb nanodots on top.  A lateral lattice shift results in not only the appearance of domain boundary, but also the transformation to rectangular $(3\sqrt{3}\times3)R30^{o}$ unit cell, which are in agreement with the STM simulations. With high spatial and energy resolution, tunnelling conductance  ($\mathrm{d}I/\mathrm{d}U$) spectra have resolved an anisotropic superconducting gap $\Delta_{NiPb} \approx 1.0$ meV in the Ni-Pb surface alloy, smaller than the $\Delta_{Pb} \approx 1.3$ meV with a typical "U" shape on the Pb(111). In addition, an enhanced coherence peak height can be explained by the higher density of states at Fermi energy ($D(E_{F})$) of the Ni-Pb surface alloy. According to temperature and spatial dependencies of $\Delta_{NiPb}$, the $T_{c} \approx 7.1$ K and a short decay length $\L_{d} \approx $ 3.55 nm have been further revealed, both are supportive results for the proximitized superconductivity in atomic-thick Ni-Pb alloy compound induced by the Pb(111) substrate.

\section{Results and discussion}

Fig. 1(a) is an overview STM topography of as-prepared sample surface after depositing Ni onto Pb(111) at elevated temperature about 400 K. Two white arrows in Fig. 1(a) indicates typical surface features of nano-cavities formed by embedded Ar-ions after sputtering and annealing process on Pb(111)\cite{MSchmid,MMuller}. Fig. 1(b) shows the corresponding atomic resolution image on Pb(111) surface and a lattice constant of 3.47 \AA\, has been extracted. In the Fig. 1(c), the line profile taken from the white dashed line across Pb(111) and ordered superstructure with a surface coverage of about 0.5 ML in Fig. 1(a) reveals an apparent height difference of 1.33 \AA\,, which is smaller than a single step height either 2.86 \AA\, of Pb(111) or 2.03 \AA\, of Ni(111). In addition, the apparent height difference at various bias voltages has also been examined, displaying a rather small variation within $\pm$ 0.1 \AA\, in the bias range of $\pm$ 1.0 V (see Supplementary Fig. S1 for details). These results imply an intermixing of Ni and Pb in the ordered superstructure on a basis of the lower apparent height than those expected from simple hard-sphere arguments\cite{MJHarrison,DFLi}. Note that the high temperature growth of  $\sqrt{3}\times\sqrt{3} R30^{0}$ Ni$_{2}$Pb$_{1}$ surface alloy has been reported by depositing Pb onto Ni(111) in previous studies\cite{JLibra,YXZhang}.

According to the atomic resolution of Pb(111) and the embedded Ar-filled nano-cavities, we are allowed to determine the lattice constant and the directions of high symmetry crystalline axes, respectively. As shown in Fig. 1(d) and (e), the ordered superstructure is composed of regular nanodot arrays with two types of arrangements, one shows rectangular lattice $(3\sqrt{3}\times3)R30^{0}$, i.e., with respect to $(1\times1)$ lattice of Pb(111), accompanied by a twofold rotation symmetry and the other is hexagonal lattice $(3\sqrt{3}\times3\sqrt{3})R30^{0}$ with a threefold rotation symmetry. Note that such ordered superstructure only appears by depositing Ni on Pb(111) at elevated temperature ranged from 400 to 550 K, and these two arrangements always coexist implying for similar energy barrier required for structural stability and growth formation. It is also noted that a lateral lattice shift creates a domain boundary in the rectangular nanodot array as indicated by yellow dashed line in Fig. 1(d). Besides, as marked by yellow arrow in Fig. 1(e), one missing dot-like feature reveals more details of underlying structure, which is an indicative of at least a bilayer stacking of ordered superstructure. Moreover, a lateral lattice shift enables the structural transformation from hexagonal to rectangular nanodot array, which can be clearly observed from the region at bottom left of Fig. 1(e).

\begin{figure}[h]
  \includegraphics[width=8.5cm]{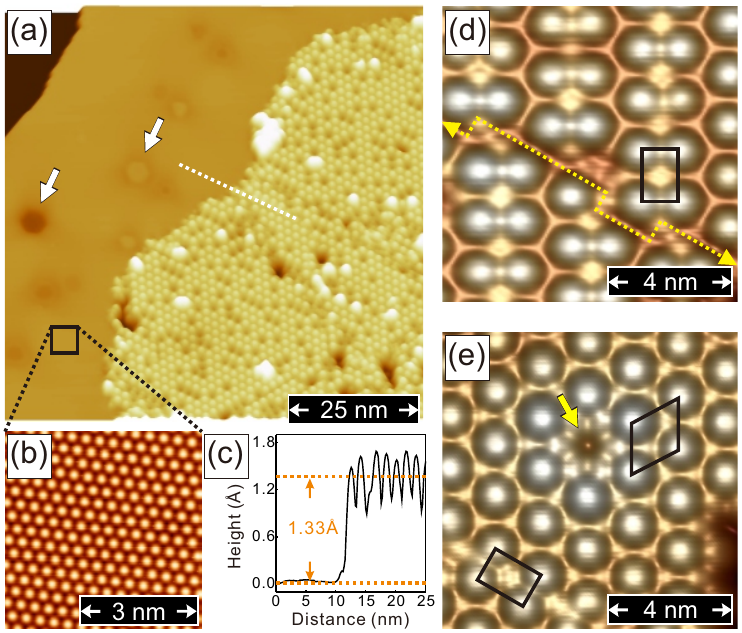}
  \caption{(a)\, Overview of STM constant-current topography of atomic-thick Ni-Pb surface alloy with an effective coverage of 0.5 ML after depositing Ni onto Pb(111) surface at 400 K. Two white arrows indicate surface features of Ar-induced nano-cavities after sputtering and annealing process on Pb(111) ($U_{b} = +1.0$\,V, $I_{t} = 0.4$\,nA). (b)\,Atomic resolution image of Pb(111) with a lattice constant of 3.47 \AA\, ($U_{b} = +10$\,mV, $I_{t} = 1.0$\,nA). (c) The apparent height of Ni-Pb alloy (white dashed line in (a)) is about 1.33 \AA\,. (d)\,Zoom-in image of rectangular $(3\sqrt{3}\times3)R30^{0}$ Ni-Pb alloy ($U_{b} = +1.0$\,V, $I_{t} = 0.4$\,nA). A domain boundary (yellow dashed arrow line) can be created by shifting a half unit-cell lattice. (e)\,Zoom-in image of hexagonal $(3\sqrt{3}\times3\sqrt{3})R30^{0}$ Ni-Pb surface alloy ($U_{b} = +1.0$\,V, $I_{t} = 0.4$\,nA). A missing nanodot as indicated by the yellow arrow suggests more details of lattice structure underneath. Transforming from hexagonal to rectangular array by a lateral lattice shift has been clearly observed at the left bottom of (e).}
  \label{fgr:Topo}
\end{figure}

When depositing Ni on Pb(111) at a higher substrate temperature of 500 K, we have found that there are more missing dot-like features on this ordered superstructure, resulting in detailed configuration of underlying layer exposed to surface as directly visualized in Fig. 2(a). Note that such ordered superstructure is restricted to the first surface layer only because we did not observe any other well-defined surface structures after depositing Ni at even higher substrate temperature, but merely big 3D clusters with a height of several nanometers. On top of that, although the ordered superstructure can form and distribute over the Pb(111) surface, it shows no signs of growing across the step edge, indicating the absence of more than a single layer of ordered superstructure at lower Pb(111) terrace. These two observations offer important clues that the formation of ordered superstructure only limits to the outermost atomic layer on surface.

Despite the existence of two types of nanodot arrangements on the ordered superstructure, the dot-like protrusions all have the uniform size and the indentical shape, a single nanodot thus indicates a nanocluster with the same constituent element, i.e., either Ni or Pb atoms. According to the first-principle adsorption energy calculations, we have been aware of that Ni atoms prefer to penetrate into Pb(111) to develop subsitutional surface alloy at elevated growth temperature. On the other hand, due to a lower cohesive energy, the Pb atoms would like to stay on top of surface giving rise to bright protrusions of nanodot arrays. Along with these understandings, at the bottom of Fig. 2(a), the side view of proposed structure model represents a bilayer stacking of the ordered superstructure consisting of a monolayer Ni-Pb alloy phase with Pb nanodots sitting on top.

Fig. 2(b) shows the zoom-in STM image of monolayer Ni-Pb surface alloy with partial missing Pb nanodots. After several attempts have been made to find the stable structure in the first-principle calculations, the monolayer Ni-Pb binary alloy with a composition of Ni$_{15}$Pb$_{12}$ has been derived from the structure model shown at the bottom of Fig. 2(b), where the black rhombus refers to the unit cell of hexagram-like network mimicked by Pb atoms (blue dots). Each center of coordinated hexagram-like lattice represents the region mainly occupied by Ni atoms with dilute Pb atom density, where the clustering of Pb atoms on top of Ni atoms becomes more energetically preferred. In the zoom-in STM image of Fig. 2(c), the monolayer Ni-Pb alloy phase has been fully covered by Pb nanodots, such that a bilayer stacking can be deduced by combining the Ni$_{15}$Pb$_{12}$ in Fig. 2(b) with additional six Pb atoms resided above (see Supplementary Fig.S2 for details). Therefore, this corresponds to an equivalent composition of Ni$_{15}$Pb$_{18}$ surface alloy that is in stoichiometric analogy used to assign the bilayer Fe-Sb surface alloy in the Fe/Sb(111) as reported by Yu \textit{et al.}\cite{YYu} Furthermore, transforming from hexagonal to rectangular arrangement of Pb nanodots can also be deduced from the rhombus to the rectangular unit cell as shown in the structure model of Fig. 2(c) (lower panel).

Besides the stable structure derived from the first-principle calculations, we have also performed the STM simulations to compare with the topography resolved experimentally. As shown in Fig. 2(d), based on higher charge density and larger atomic size calculated in the Pb atoms than in the Ni atoms, the simulated STM images have reproduced the hexagram-like structure as well as the $(3\sqrt{3}\times3\sqrt{3})R30^{o}$ to $(3\sqrt{3}\times3)R30^{o}$ lattice transformation, which are in good agreement with the experimental results in Fig. 2(b) and (c), respectively. Due to a significant amount of charge transfer from Pb to Ni, we would also like to denote that the delocalization of \textit{d} electrons turns out leading to the vanishing magnetic moment in the Ni atoms of the Ni-Pb surface alloy and similar tendency has also been reported on the weakening of ferromagnetism when Ni-Pb surface alloy formed on Ni(111) surface\cite{JLibra,YXZhang}.

\begin{figure}[!ht]
  \includegraphics[width=8.5cm]{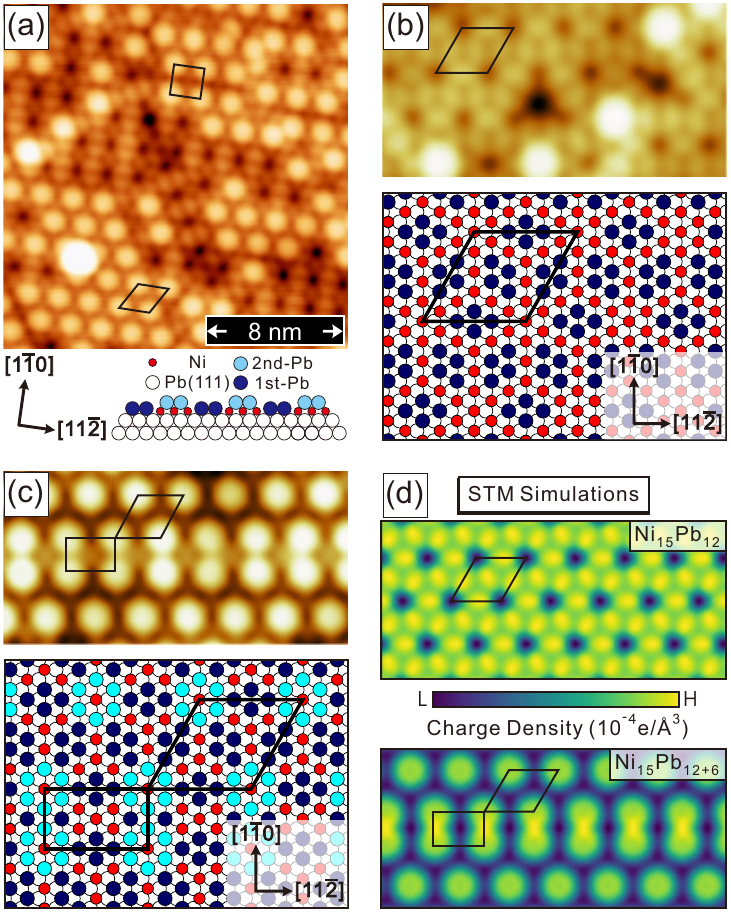}
  \caption{(a)\, Top: STM topography of Ni-Pb surface alloy with several missing Pb nanodots. Bottom: side view of structure model with a bilayer stacking of Pb nanodots sitting on top of monolayer Ni-Pb alloy phase ($U_{b} = +1.0$\,V, $I_{t} = 1.0$\,nA). (b)\, Zoom-in image of monolayer Ni-Pb alloy phase, black rhombus indicates the unit cell of a coordinated hexagram-like network formed by Ni$_{15}$Pb$_{12}$ binary alloy as proposed in structure model at the bottom ($U_{b} = +1.0$\,V, $I_{t} = 1.0$\,nA). (c)\, Zoom-in image of the Ni-Pb surface alloy fully covered with Pb nanodots on top, such bilayer stacking turns out having an equivalent composition of Ni$_{15}$Pb$_{12+6}$. The hexagonal to rectangular array of Pb nanodots can be transformed from rhombus to rectangular unit cell by introducing a lateral lattice shift to the structure model at the bottom ($U_{b} = +1.0$\,V, $I_{t} = 1.0$\,nA). (d) The STM simulations based on the structure models reproduce the consistent features as resovled experimentally in topography images (b) and (c), respectively.}
  \label{fgr:STS}
\end{figure}

Since the tunnelling spectroscopy can provide high spatial and energy resolution, the point conductance spectra have been carried out to access the proximity-induced superconductivity in the Ni-Pb surface alloy grown on Pb(111). It is known that the bulk material of Pb is the conventional Bardeen–Cooper–Schrieffer (BCS)-type superconductor, the feature of superconducting gap ($\Delta$) opened at Fermi energy $E_{F}$ is therefore expected to appear in $\mathrm{d}I/\mathrm{d}U$ curve as the black line displayed in the middle panel of Fig. 3(a) when measurement temperature is below superconducting transition temperature ($T_{c}$) at 7.2 K on Pb(111)\cite{JKim,JKim1}. In addition to the bulk Pb(111) substrate, interestingly, the superconducting gap remains clearly to be observed on the Ni-Pb surface alloy as shown in Fig. 3(a) (blue line in lower panel), but with a smaller gap size and higher intensity of coherence peak. Most importantly, instead of typical "U" shape of superconducting gap as resolved on Pb(111), the anisotropic superconducting gap revealed on the Ni-Pb surface alloy is unusual for an atomic-thick bimetallic compound and has not been reported yet. We denote that the $\mathrm{d}I/\mathrm{d}U$ spectrum taken from the Ni-Pb surface alloy has been overlapped with the one from Pb(111) for a direct comparison, and we have also performed a control experiment by quenching the superconducting gap through external magnetic filed to identify the type-I superconductivity and exclude the possible tip-induced artifacts (see Supplementary Fig.S3 for details). 

In order to have a quantitative comparison, the normalized $\mathrm{d}I/\mathrm{d}U$ curve of Pb(111) have been fitted to the BCS-like density of states (DOS), and the corresponding values of $\Delta_{Pb} \approx 1.3$ meV has been obtained (red line in middle panel of Fig. 3(a)). However, in the case of Ni-Pb surface alloy, we can not simply apply the BCS-fitting because of either the incorrect gap size or the wrong energy position of coherence peak (see Supplementary Fig.S4 for details). Considering the reduced and anisotropic superconducting gap of Ni-Pb surface alloy, it suggests us an attempt of using the \textit{d}-wave fitting with typical "V" shape of superconducting gap, but we still fail to restore these features in the $\mathrm{d}I/\mathrm{d}U$ spectrum (see Supplementary Fig.S4 for details). Rather than the \textit{d}-wave fitting, interestingly, we have found the anisotropic gap fitting from anisotropic \textit{s}-wave model, i.e., nonuniform gap opening in different momentum directions on the Fermi contour, is necessary to reproduce the $\mathrm{d}I/\mathrm{d}U$ curve as the red line shown in lower panel of Fig. 3(a). Although we have noticed that the anisotropic gap fitting deviates slightly to fully recover the shoulder of coherence peak, it still fairly agrees with experimental $\mathrm{d}I/\mathrm{d}U$ curve of Ni-Pb surface alloy, especially in the critical energy range covering the superconducting gap, enabling us to obtain an appropriate value of $\Delta_{NiPb} \approx 1.0$ meV (see Supplementary Fig.S4 for details). We denote that the anisotropic superconductivity appeared in the atomic-thick surface alloy via proximity effect remains unprecedented, and it was exclusively reported on the bulk compound materials before\cite{KMaki,TWatanabe,WHJiao,ZDu,RProzorov,SYasuzuka}.

After revealing the anisotropic superconducting gap, we have further found an enhancement of coherence peak intensity and it's robustness against the structural variations in the monolayer Ni-Pb binary alloy. As shown in Fig. 3(b), the point conductance spectra have been taken at different atomic sites on the Ni-Pb surface alloy, including bridge (red), on top (green) and vacancy from missing single Pb nanodot (blue). The enhanced coherence peak height as compared to Pb(111) has always been observed and does not correlate with the structural changes. This finding, in particular the coherence height, is on strong contrary to the proximity-induced superconductivity in monolayer MoS$_{2}$ reported recently by Trainer \textit{et al.}, whereas the intensity of coherence peak is spatially modulated and has a registry with the moir\'e pattern\cite{DJTrainer}.

According to the angle-resolved photoemission spectroscopy (ARPES) studies done by Libra \textit{et al.}, they have pointed out the noticeable changes of DOS nearby the Fermi level when the Ni-Pb surface alloy formed on the Ni(111)\cite{JLibra}. Inspired by their work, we have further performed a detailed numerical calculation on the basis of Bogoliubov-de Gennes (BdG) Hamiltonian to gain an insight into the enhanced coherence peak height in the superconducting Ni-Pb surface alloy on Pb(111) (see Supplementary Note I for details). From the DFT calculations and experimental spectroscopy measurements, a higher $D(E_{F})$ in the normal state of Ni-Pb surface alloy as compared to Pb(111) has been considered in the numerical model, and an enhancement of coherence peak height can therefore be obtained in the $\mathrm{d}I/\mathrm{d}U$ spectrum of superconducting Ni-Pb surface alloy by solving the BdG equations. It's noted that such higher $D(E_{F})$ is originated from the intrinsic electronic properties of Ni-Pb surface alloy grown on the Pb(111), which could explain why the enhanced coherence peak height is not sensitive to the local structural perturbations.

To understand the temperature dependence of proximitized superconducting Ni-Pb surface alloy, the temperature-dependent $\mathrm{d}I/\mathrm{d}U$ spectra have been recorded to obtain the temperature-dependent superconducting gap ($\Delta(T)$) and the transition temperature $T_{c}$. Fig. 3(c) represents a series of $\mathrm{d}I/\mathrm{d}U$ curves measured on the Pb(111) (black) and the Ni-Pb surface alloy (blue) at various temperatures. With an increase of temperature, both $\Delta_{Pb}$ and $\Delta_{NiPb}$ continue to decrease and their corresponding values as a function of temperature have been summarized in Fig. 3(d) (black and blue dots). They both vanish at about 7.1 K, i.e., in line with the $T_{c} \approx 7.2$ K for the bulk Pb, and thus yield to the same $T_{c}$ that roots for the proximity-induced superconductivity in the Ni-Pb surface alloy. According to the $2\Delta_{Pb}(0)/k_{B}T_{c} \approx 4.39 $, we can identify the bulk Pb as a strong-coupling superconductor that is on the contrary to the Ni-Pb surface alloy with $2\Delta_{NiPb}(0)/k_{B}T_{c} \approx 3.37 $ close to the weak-coupling limit ($\approx 3.52$) under the framework of BCS theory. However, by using generic BCS gap equation\cite{JBardeen, RProzorov1}, not only the $\Delta_{NiPb}(T)$, but also the $\Delta_{Pb}(T)$ can be fitted reasonably well (black and blue lines in Fig. 3(d)), which justifies the good applicability of universal curve $\Delta(T)$ in the most of cases, even with Pb\cite{MTinkham, SBose}. 

\begin{figure}[h]
  \includegraphics[width=8.5cm]{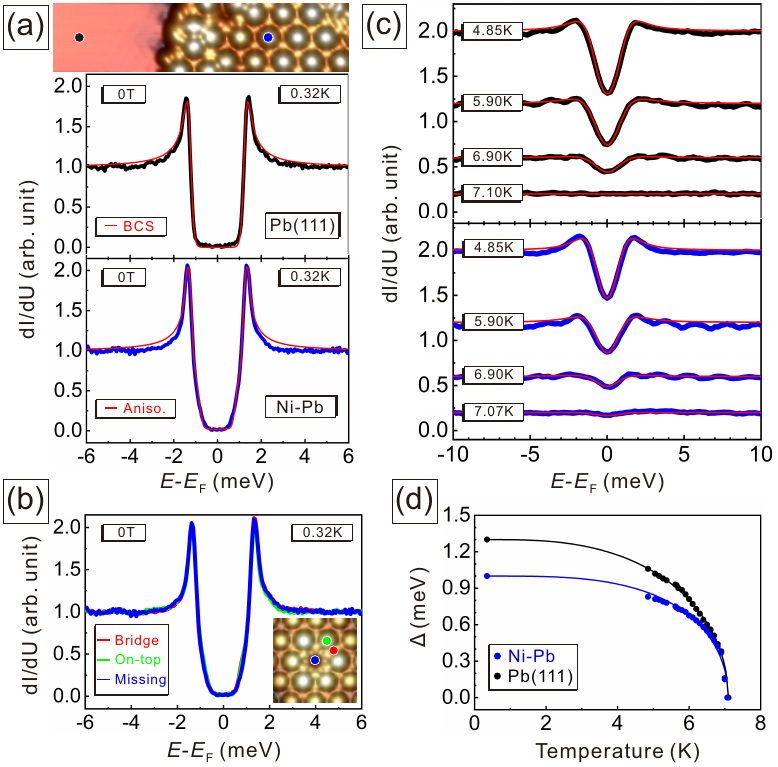}
  \caption{(a)\, Point conductance spectra taken on the Pb(111) and the Ni-Pb surface alloy as indicated by black and blue points in topography at top. According to the BCS and anisotropic gap fittings (red lines), the $\Delta_{Pb} \approx 1.3$ meV and $\Delta_{NiPb} \approx 1.0$ meV have been obtained, respectively. (b)\, Point conductance spectra taken at different atomic sites of the Ni-Pb surface alloy, including bridge (red), on top (green) and vacancy from missing single Pb nanodot (blue) as the topography shown in the inset. The robustness of proximity-induced superconducting state of the Ni-Pb surface alloy can be referred to the absence of substantial influences on resultant $\mathrm{d}I/\mathrm{d}U$ curves against the local structural perturbations. (c)\, $\mathrm{d}I/\mathrm{d}U$ curves as a function of temperature on the Pb(111) (black) and the Ni-Pb surface alloy (blue). (d)\, Temperature dependent $\Delta_{Pb}$ and $\Delta_{NiPb}$ reveals the identical $T_{c}$ about 7.1 K as extracted from universal BCS gap equation. (stabilization parameters: $U_{b} = +10$\,mV, $I_{t} = 1.0$\,nA for all $\mathrm{d}I/\mathrm{d}U$ curves)}
  \label{fgr:QPI}
\end{figure}

In addition to the point conductance curve, as shown in Fig. 4(a), the line spectroscopy taken point-by-point from the $\mathrm{d}I/\mathrm{d}U$ curves along the black dashed line in topography (top panel) has been performed to map out the spatial dispersion of $\Delta_{Pb}$ and $\Delta_{NiPb}$ in the proximity region. The $\mathrm{d}I/\mathrm{d}U$ spectra as a function of distance measured in resolution of 4 \AA\, has been presented in the Fig. 4(a) (bottom panel) and the Fig. 4(b), respectively, showing not only the evolution form the isotropic $\Delta_{Pb}$ into the anisotropic $\Delta_{NiPb}$, but also the enhancement of coherence peak height. In order to uncover the role of proximity effect in the superconducting Ni-Pb surface alloy, the superconductor ($\Delta_{Pb}$) to superconductor ($\Delta_{NiPb}$) 1D junction has been approximated in the self-consistent Usadel model (see Supplementary Note II for details). Note that the 1D diffusive model of Usadel theory has been successfully applied to explain the proximity effect in the induced superconductivity of striped incommensurate (SIC) Pb monolayer grown on Si(111)\cite{TZhang, JKim1,VCherkez}. Interestingly, the sharp transition from $\Delta_{Pb}$ to $\Delta_{NiPb}$ has nearly been replicated under the 1D assumption as shown in Fig. 4(c), although the anisotropic shape of $\Delta_{NiPb}$ and enhanced coherence peak height are not considered in the Usadel theory. The spatial dispersion of $\Delta$ (black dots) extracted from Fig. 4(b) with a short decay length $L_{d} \approx $ 3.55 nm (gray shaded area) have been summarized in the Fig. 4(d). Such short decay length can not be fully recovered by the self-consistent fitting (red dots) in the 1D approximation of Usadel model as shown in Fig. 4(d), indicating the contribution of proximity effect from the underneath Pb(111) substrate needs to be taken into account for the superconducting Ni-Pb surface alloy. Moreover, the reflectivity coefficient $r = 0.02$ has been determined from the discontinuity of $\Delta$ at the interface as shown in Fig. 4(d), referring to an interfacial transparency for the penetration of Cooper pairs from the Pb(111) substrate that turns the Ni-Pb surface alloy into superconducting (see Supplementary Note II for details). In Fig. 4(e), we have further analyzed the coherence peak asymmetry, i.e., the intensity difference between coherence peak and normalized background divided by their sum, to quantify about  4 $\%$ increase of coherence peak height (black dots) in the Ni-Pb surface alloy. On the contrary, such enhancement of coherence peak height is absent from the typical proximity effect considered in the diffusive model of Usadel theory, whereas the coherence peak asymmetry (red dots) in the Ni-Pb surface alloy asymptotically decreases instead (see Supplementary Note II for details).

\begin{figure}[!ht]
  \includegraphics[width=8.5cm]{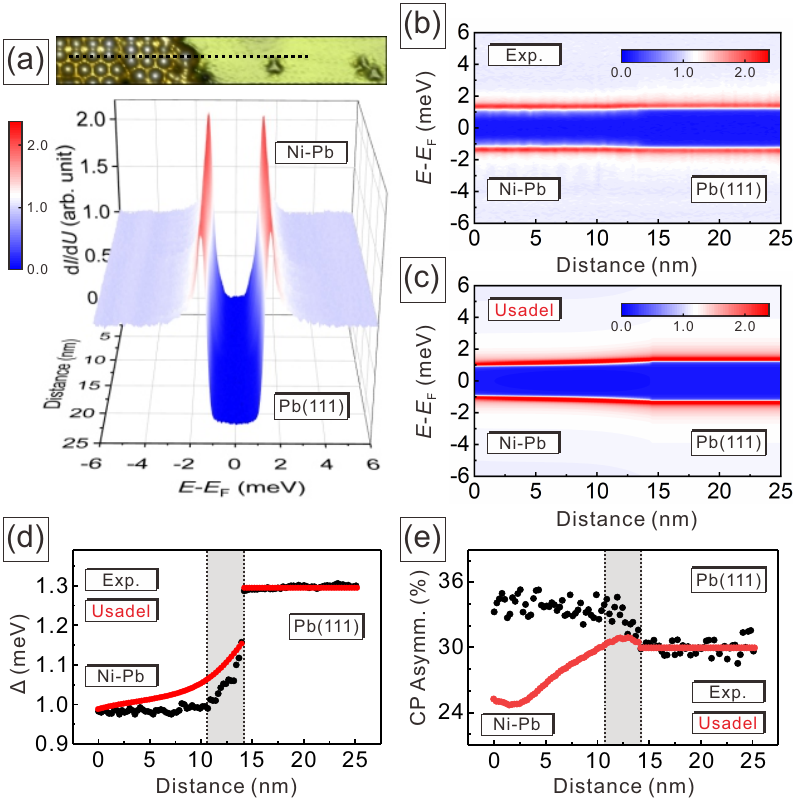}
  \caption{(a) Spatial dependent $\mathrm{d}I/\mathrm{d}U$ spectra measured along the black dashed line in topography (top panel) represents the reduced but anisotropic superconducting gap in the Ni-Pb surface alloy (bottom panel). (b) The sharp transition from the discontinuity of superconducting gap and the enhanced intensity of coherence peak height have been clearly resolved. (c) The self-consistent Usadel model essentially captures the quick drop of superconducting gap in (b). (d) The experimental profile of $\Delta$ (black dots) with a short decay length $L_{d} \approx $ 2.75 nm (gray shaded area) can not be fully simulated from the self-consistent fitting in the 1D approximation of Usadel model. (e) The experimental observation of enhanced coherence peak height (black dots) corresponds to about 4 $\%$ increase of coherence peak asymmetry, which is contrast to the diffusive proximity effect generally considered in the Usadel model (red dots). (stabilization parameters: $U_{b} = +10$\,mV, $I_{t} = 1.0$\,nA for all $\mathrm{d}I/\mathrm{d}U$ curves) }
  \label{fgr:QPI}
\end{figure}

While the proximity effect is known to stabilize the local superconducting order within a finite distance into the normal metal, we can expect a smaller superconducting gap but with the "U" shape in the Ni-Pb surface alloy, the consistent results that can be derived from the diffusive Usadel model. However, the proximity effect along is not sufficient to explain the anisotropic superconductivity as resolved in the Ni-Pb surface alloy, i.e., the anisotropy occurs in opening the superconducting gap at different momentum directions on the Fermi surface\cite{KMaki,TWatanabe,WHJiao,ZDu,RProzorov,SYasuzuka}, which is distinct from the proximity-induced isotropic superconducting gap by the conventional BCS-type Pb(111) substrate underneath. Nevertheless, in perspective of the fundamental BCS theory\cite{JBardeen, RProzorov1}, we are aware of that the superconducting gap size $\Delta$ can be related to the Debye frequency ($\omega_{D}$) and dimensionless coupling constant $\lambda = D(E_{F})V_{ep}$, i.e., $V_{ep}$ is electron-phonon pairing potential, which could contribute to the anisotropy in opening the momentum dependent superconducting gap on the Fermi surface.

From the growth studies, the Ni-Pb surface alloy can only be fabricated on Pb(111) at elevated temperature ranged from 400 to 550 K below the melting temperature of Pb ($T_{m(Pb)}$), meaning the $\omega_{D(NiPb)}$ could likely be lower than the $\omega_{D(Pb)}$, since $T_{m}$ is proportional to the square of $\omega_{D}$ according to the Lindemann's criterion\cite{FLindemann, SAKhrapak}. Therefore, the $\Delta_{NiPb}$ can be smaller than the $\Delta_{Pb}$ because the $\omega_{D(NiPb)}$ is lower than the $\omega_{D(Pb)}$, if the same coupling constant $\lambda$ has been assumed. However, since the $\omega_{D}$ is the highest frequency the lattice phonon can be excited, the comparison between $\omega_{D(NiPb)}$ and $\omega_{D(Pb)}$ can only explain the reduction value of $\Delta_{NiPb}$, but not the anisotropy in the superconducting gap opening.

On the other hand, we can compare the $\lambda_{NiPb}$ with $\lambda_{Pb}$, and deduce the $V_{ep(NiPb)}$ weaker than the $V_{ep(Pb)}$ from the results of the smaller $\Delta_{NiPb}$ and the higher $D(E_{F})$ in the Ni-Pb surface alloy. In light of the stronger $V_{ep}$ achieved by increasing surface strain in various graphene-based superconductors\cite{CSi, YGe, GLi}, one might be able to associate the weaker $V_{ep(NiPb)}$ with the reduction of surface tensile stress in the formation of Ni-Pb surface alloy, thus leading to a smaller $\Delta_{NiPb}$ resolved experimentally. We denote that the formation of surface alloy (or surface reconstruction in general) has been recognized more energetically favorable regarding to a relief of surface tensile stress as reported on the $(\sqrt{3}\times\sqrt{3})R30^{o}$ Ni-Pb surface alloy (or well-known Au(111) herring-bone surface reconstruction) in previous studies\cite{DPWoodruff, MJHarrison}. In addition to the $V_{ep(NiPb)}$ weaker than the $V_{ep(Pb)}$, most importantly, the $V_{ep(NiPb)}$ could be anisotropic arising from the inversion symmetry breaking and structural inhomogeneity of coexisting hexagonal $(3\sqrt{3}\times3\sqrt{3})R30^{o}$ and rectangular $(3\sqrt{3}\times3)R30^{o}$ lattices in the Ni-Pb surface alloy. Therefore, the anisotropic $V_{ep(NiPb)}$ could lead to an anisotropy in the superconducting gap function, and thus plays an essential role in the emergence of anisotropic superconductivity from the Ni-Pb surface alloy.

\section{Conclusions}

In conclusion, we have systematically investigated the proximity-induced anisotropic superconductivity in monolayer Ni-Pb binary alloy. Through the high temperature deposition of Ni onto the Pb(111),  the $(3\sqrt{3}\times3\sqrt{3})R30^{o}$ Ni-Pb surface alloy has been grown with uniform Pb nanodots stacking on top. The domain boundary and the $(3\sqrt{3}\times3)R30^{o}$ lattice transformation have been deduced from a lateral shift in the unit cell, which are in line with the simulated STM results. By utilizing tunneling spectroscopy with high spatial and energy resolution, we have resolved a reduced but anisotropic superconducting gap $\Delta_{NiPb} \approx 1.0$ meV on the Ni-Pb surface alloy, in strong contrast to an isotropic superconducting gap $\Delta_{Pb} \approx 1.3$ meV on the Pb(111). In addition, the higher $D(E_{F})$ of the Ni-Pb surface alloy contributes to an enhancement of coherence peak height, resulting in about 4 $\%$ asymmetry larger than the Pb(111). According to temperature and spatial dependencies of $\Delta_{NiPb}$, the $T_{c} \approx 7.1$ K and a short decay length $L_{d} \approx $ 3.55 nm have been revealed, further supporting the Pb(111) substrate as the main driving force for developing superconductivity via proximity effect. In light of anisotropic superconductivity emerged from the Ni-Pb surface alloy, our results demonstrate an important capability of tailoring proximity-induced superconductivity through fabricating the single-atomic-layer bimetallic compound, whereas the immiscible bulk counterpart does not even exist.

\section{Experimental Sections}
\subsection{Experimental}
The Ni/Pb(111)were prepared in an ultrahigh vacuum (UHV) chamber with the base pressure below $p \leq 2 \times 10^{-10}$\,mbar. The clean Pb(111) surface was first prepared by cycles of Ar$^{+}$ ion sputtering with an ion energy of 500\,eV at room temperature and subsequent annealing up to 600\,K. The Ni source with purity of 99.999\,\% (Goodfellow) was e-beam sublimated onto Pb(111) surface at elevated temperature ranged from 400 to 550\,K at which the well-ordered and extended Ni-Pb surface alloy with $(3\sqrt{3}\times3\sqrt{3})R30^{o}$ and $(3\sqrt{3}\times3)R30^{o}$ lattice structures can be grown. 
After preparation, the sample was immediately transferred into a ultralow-temperature scanning tunneling microscope (LT-STM) from Unisoku Co. Ltd. (operation temperature $T \approx 0.32$\,K). The topography images were obtained from the constant-current mode with the bias voltage $U$ applied to the sample. For scanning tunneling spectroscopy (STS) measurements, a small bias voltage modulation was added to $U$ (frequency $\nu = 3991$\,Hz), such that tunneling differential conductance $\mathrm{d}I/\mathrm{d}U$ spectra can be acquired by detecting the first harmonic signal by means of a lock-in amplifier.


\begin{acknowledgement}

Y.H.L. and C.H.H. and I.J. contributed equally to this work. D.S.L. and P.J.H. acknowledge support from the competitive research funding from National Tsing Hua University, Ministry of Science and Technology of Taiwan under Grants No. MOST-110-2636-M-007-006 and MOST-110-2124-M-A49-008-MY3, and center for quantum technology from the featured areas research center program within the framework of the higher education sprout project by the Ministry of Education (MOE) in Taiwan.

\end{acknowledgement}

\begin{suppinfo}

Bias dependent apparent height of Ni-Pb surface alloy on Pb(111), atomic resolution in proximity between Ni-Pb surface alloy and Pb(111), superposition and magnetic field dependence of dI/dU spectra, comparison of BCS-fitting, d-wave and anisotropic gap fittings, simulation of enhanced coherence peak height by BdG model and self-consistent fitting by the diffusive model of Usadel theory can be found in the Supporting Information.

\end{suppinfo}

\bibliography{Ni_Pb111_1stSubm_Ref_20210916}

\providecommand{\latin}[1]{#1}
\makeatletter
\providecommand{\doi}
  {\begingroup\let\do\@makeother\dospecials
  \catcode`\{=1 \catcode`\}=2 \doi@aux}
\providecommand{\doi@aux}[1]{\endgroup\texttt{#1}}
\makeatother
\providecommand*\mcitethebibliography{\thebibliography}
\csname @ifundefined\endcsname{endmcitethebibliography}
  {\let\endmcitethebibliography\endthebibliography}{}
\begin{mcitethebibliography}{59}
\providecommand*\natexlab[1]{#1}
\providecommand*\mciteSetBstSublistMode[1]{}
\providecommand*\mciteSetBstMaxWidthForm[2]{}
\providecommand*\mciteBstWouldAddEndPuncttrue
  {\def\EndOfBibitem{\unskip.}}
\providecommand*\mciteBstWouldAddEndPunctfalse
  {\let\EndOfBibitem\relax}
\providecommand*\mciteSetBstMidEndSepPunct[3]{}
\providecommand*\mciteSetBstSublistLabelBeginEnd[3]{}
\providecommand*\EndOfBibitem{}
\mciteSetBstSublistMode{f}
\mciteSetBstMaxWidthForm{subitem}{(\alph{mcitesubitemcount})}
\mciteSetBstSublistLabelBeginEnd
  {\mcitemaxwidthsubitemform\space}
  {\relax}
  {\relax}

\bibitem[de~Gennes(1964)]{PGdeGennes}
de~Gennes,~P.~G. \emph{Rev.\ Mod.\ Phys.} \textbf{1964}, \emph{36}, 225\relax
\mciteBstWouldAddEndPuncttrue
\mciteSetBstMidEndSepPunct{\mcitedefaultmidpunct}
{\mcitedefaultendpunct}{\mcitedefaultseppunct}\relax
\EndOfBibitem
\bibitem[Hauser \latin{et~al.}(1966)Hauser, Theuerer, and Werthamer]{JJHauser}
Hauser,~J.~J.; Theuerer,~H.~C.; Werthamer,~N.~R. \emph{Phys.\ Rev.\ B}
  \textbf{1966}, \emph{142}, 118\relax
\mciteBstWouldAddEndPuncttrue
\mciteSetBstMidEndSepPunct{\mcitedefaultmidpunct}
{\mcitedefaultendpunct}{\mcitedefaultseppunct}\relax
\EndOfBibitem
\bibitem[Beenakker(1997)]{CWJBeenakker}
Beenakker,~C. W.~J. \emph{Rev.\ Mod.\ Phys.} \textbf{1997}, \emph{69},
  731\relax
\mciteBstWouldAddEndPuncttrue
\mciteSetBstMidEndSepPunct{\mcitedefaultmidpunct}
{\mcitedefaultendpunct}{\mcitedefaultseppunct}\relax
\EndOfBibitem
\bibitem[Lambert and Raimondi(1998)Lambert, and Raimondi]{CLambert}
Lambert,~C.; Raimondi,~R. \emph{J.\ Phys.\ Condens.\ Matter.} \textbf{1998},
  \emph{10}, 901\relax
\mciteBstWouldAddEndPuncttrue
\mciteSetBstMidEndSepPunct{\mcitedefaultmidpunct}
{\mcitedefaultendpunct}{\mcitedefaultseppunct}\relax
\EndOfBibitem
\bibitem[Buzdin(2005)]{AIBuzdin}
Buzdin,~A.~I. \emph{Rev.\ Mod.\ Phys.} \textbf{2005}, \emph{77}, 935\relax
\mciteBstWouldAddEndPuncttrue
\mciteSetBstMidEndSepPunct{\mcitedefaultmidpunct}
{\mcitedefaultendpunct}{\mcitedefaultseppunct}\relax
\EndOfBibitem
\bibitem[F.~S.~Bergeret and Efetov(2005)F.~S.~Bergeret, and Efetov]{FSBergeret}
F.~S.~Bergeret,~A. F.~V.; Efetov,~K.~B. \emph{Rev.\ Mod.\ Phys.} \textbf{2005},
  \emph{77}, 1321\relax
\mciteBstWouldAddEndPuncttrue
\mciteSetBstMidEndSepPunct{\mcitedefaultmidpunct}
{\mcitedefaultendpunct}{\mcitedefaultseppunct}\relax
\EndOfBibitem
\bibitem[Fu and Kane(2008)Fu, and Kane]{LFu}
Fu,~L.; Kane,~C.~L. \emph{Phys.\ Rev.\ Lett.} \textbf{2008}, \emph{100},
  096407\relax
\mciteBstWouldAddEndPuncttrue
\mciteSetBstMidEndSepPunct{\mcitedefaultmidpunct}
{\mcitedefaultendpunct}{\mcitedefaultseppunct}\relax
\EndOfBibitem
\bibitem[Qi \latin{et~al.}(2009)Qi, Hughes, Raghu, and Zhang]{XLQi}
Qi,~X.~L.; Hughes,~T.~L.; Raghu,~S.; Zhang,~S.-C. \emph{Phys.\ Rev.\ Lett.}
  \textbf{2009}, \emph{102}, 187001\relax
\mciteBstWouldAddEndPuncttrue
\mciteSetBstMidEndSepPunct{\mcitedefaultmidpunct}
{\mcitedefaultendpunct}{\mcitedefaultseppunct}\relax
\EndOfBibitem
\bibitem[Williams \latin{et~al.}(2012)Williams, Bestwick, Gallagher, Hong, Cui,
  Bleich, Analytis, Fisher, and Goldhaber-Gordon]{JRWilliams}
Williams,~J.~R.; Bestwick,~A.~J.; Gallagher,~P.; Hong,~S.~S.; Cui,~Y.;
  Bleich,~A.~S.; Analytis,~J.~G.; Fisher,~I.~R.; Goldhaber-Gordon,~D.
  \emph{Phys.\ Rev.\ Lett.} \textbf{2012}, \emph{109}, 056803\relax
\mciteBstWouldAddEndPuncttrue
\mciteSetBstMidEndSepPunct{\mcitedefaultmidpunct}
{\mcitedefaultendpunct}{\mcitedefaultseppunct}\relax
\EndOfBibitem
\bibitem[Veldhorst \latin{et~al.}(2012)Veldhorst, Snelder, Hoek, Gang, Guduru,
  X.~L.~Wang, der wiel, Golubov, Hilgenkamp, and Bririkman]{MVeldhorst}
Veldhorst,~M.; Snelder,~M.; Hoek,~M.; Gang,~T.; Guduru,~V.~K.;
  X.~L.~Wang,~U.~Z.; der wiel,~W. G.~V.; Golubov,~A.~A.; Hilgenkamp,~H.;
  Bririkman,~A. \emph{Nat.\ Mater.} \textbf{2012}, \emph{11}, 417\relax
\mciteBstWouldAddEndPuncttrue
\mciteSetBstMidEndSepPunct{\mcitedefaultmidpunct}
{\mcitedefaultendpunct}{\mcitedefaultseppunct}\relax
\EndOfBibitem
\bibitem[Wang \latin{et~al.}(2012)Wang, Liu, Xu, Yang, Miao, Yao, Gao, Shen,
  Ma, Xu, Liu, Zhang, Qian, Jia, and Xue]{MXWang}
Wang,~M.~X.; Liu,~C.; Xu,~J.~P.; Yang,~F.; Miao,~L.; Yao,~M.~Y.; Gao,~C.~L.;
  Shen,~C.; Ma,~X.; Xu,~Z.~A.; Liu,~Y.; Zhang,~S.~C.; Qian,~D.; Jia,~J.~F.;
  Xue,~Q.~K. \emph{Science} \textbf{2012}, \emph{52}, 336\relax
\mciteBstWouldAddEndPuncttrue
\mciteSetBstMidEndSepPunct{\mcitedefaultmidpunct}
{\mcitedefaultendpunct}{\mcitedefaultseppunct}\relax
\EndOfBibitem
\bibitem[Bergeret \latin{et~al.}(2001)Bergeret, Volkov, and
  Efetov]{FSBergeret1}
Bergeret,~F.~S.; Volkov,~A.~F.; Efetov,~K.~B. \emph{Phys.\ Rev.\ Lett.}
  \textbf{2001}, \emph{86}, 4096--4099\relax
\mciteBstWouldAddEndPuncttrue
\mciteSetBstMidEndSepPunct{\mcitedefaultmidpunct}
{\mcitedefaultendpunct}{\mcitedefaultseppunct}\relax
\EndOfBibitem
\bibitem[Kadigrobov \latin{et~al.}(2001)Kadigrobov, Shekhter, and
  Jonson]{AKadigrobov}
Kadigrobov,~A.; Shekhter,~R.~I.; Jonson,~M. \emph{Europhys.\ Lett.}
  \textbf{2001}, \emph{54}, 394--400\relax
\mciteBstWouldAddEndPuncttrue
\mciteSetBstMidEndSepPunct{\mcitedefaultmidpunct}
{\mcitedefaultendpunct}{\mcitedefaultseppunct}\relax
\EndOfBibitem
\bibitem[Eschrig \latin{et~al.}(2003)Eschrig, Kopu, Cuevas, and
  Sch\"{o}n]{MEschrig}
Eschrig,~M.; Kopu,~J.; Cuevas,~J.~C.; Sch\"{o}n,~G. \emph{Phys.\ Rev.\ Lett.}
  \textbf{2003}, \emph{90}, 137003\relax
\mciteBstWouldAddEndPuncttrue
\mciteSetBstMidEndSepPunct{\mcitedefaultmidpunct}
{\mcitedefaultendpunct}{\mcitedefaultseppunct}\relax
\EndOfBibitem
\bibitem[Keizer \latin{et~al.}(2006)Keizer, Goennenwein, Klapwijk, Miao, Xiao,
  and Gupta]{RSKeizer}
Keizer,~R.~S.; Goennenwein,~S. T.~B.; Klapwijk,~T.~M.; Miao,~G.; Xiao,~G.;
  Gupta,~A. \emph{Nature} \textbf{2006}, \emph{436}, 825--827\relax
\mciteBstWouldAddEndPuncttrue
\mciteSetBstMidEndSepPunct{\mcitedefaultmidpunct}
{\mcitedefaultendpunct}{\mcitedefaultseppunct}\relax
\EndOfBibitem
\bibitem[Yang \latin{et~al.}(2010)Yang, Yang, Takahashi, Maekawa, and
  Parkin]{HYang}
Yang,~H.; Yang,~S.~H.; Takahashi,~S.; Maekawa,~S.; Parkin,~S. S.~P. \emph{Nat.\
  Mat.} \textbf{2010}, \emph{9}, 586--593\relax
\mciteBstWouldAddEndPuncttrue
\mciteSetBstMidEndSepPunct{\mcitedefaultmidpunct}
{\mcitedefaultendpunct}{\mcitedefaultseppunct}\relax
\EndOfBibitem
\bibitem[Eschrig(2011)]{MEschrig1}
Eschrig,~M. \emph{Phys.\ Today} \textbf{2011}, \emph{64}, 4349\relax
\mciteBstWouldAddEndPuncttrue
\mciteSetBstMidEndSepPunct{\mcitedefaultmidpunct}
{\mcitedefaultendpunct}{\mcitedefaultseppunct}\relax
\EndOfBibitem
\bibitem[Xu \latin{et~al.}(2015)Xu, Wang, Liu, Ge, Yang, Liu, Xu, Guan, Gao,
  Qian, Liu, Wang, Zhang, Xue, and Jia]{JPXu}
Xu,~J.~P.; Wang,~M.~X.; Liu,~Z.~L.; Ge,~J.~F.; Yang,~X.; Liu,~C.; Xu,~Z.~A.;
  Guan,~D.; Gao,~C.~L.; Qian,~D.; Liu,~Y.; Wang,~Q.~H.; Zhang,~F.~C.;
  Xue,~Q.~K.; Jia,~J.~F. \emph{Phys.\ Rev.\ Lett.} \textbf{2015}, \emph{114},
  017001\relax
\mciteBstWouldAddEndPuncttrue
\mciteSetBstMidEndSepPunct{\mcitedefaultmidpunct}
{\mcitedefaultendpunct}{\mcitedefaultseppunct}\relax
\EndOfBibitem
\bibitem[Wang \latin{et~al.}(2018)Wang, Kong, Fan, Chen, Zhu, Liu, Cao, Sun,
  Du, Schneeloch, Zhong, Gu, Fu, Ding, and Gao]{DFWang}
Wang,~D.; Kong,~L.; Fan,~P.; Chen,~H.; Zhu,~S.; Liu,~W.; Cao,~L.; Sun,~Y.;
  Du,~S.; Schneeloch,~J.; Zhong,~R.; Gu,~G.; Fu,~L.; Ding,~H.; Gao,~H.-J.
  \emph{Science} \textbf{2018}, \emph{362}, 333--335\relax
\mciteBstWouldAddEndPuncttrue
\mciteSetBstMidEndSepPunct{\mcitedefaultmidpunct}
{\mcitedefaultendpunct}{\mcitedefaultseppunct}\relax
\EndOfBibitem
\bibitem[Machida \latin{et~al.}(2019)Machida, Sun, Pyon, Takeda, Kohsaka,
  Hanaguri, Sasagawa, and Tamegai]{TMachida}
Machida,~T.; Sun,~Y.; Pyon,~S.; Takeda,~S.; Kohsaka,~Y.; Hanaguri,~T.;
  Sasagawa,~T.; Tamegai,~T. \emph{Nat.\ Mater.} \textbf{2019}, \emph{18},
  811--815\relax
\mciteBstWouldAddEndPuncttrue
\mciteSetBstMidEndSepPunct{\mcitedefaultmidpunct}
{\mcitedefaultendpunct}{\mcitedefaultseppunct}\relax
\EndOfBibitem
\bibitem[Alicea(2012)]{JAlicea}
Alicea,~J. \emph{Rep.\ Prog.\ Phys.} \textbf{2012}, \emph{75}, 076501\relax
\mciteBstWouldAddEndPuncttrue
\mciteSetBstMidEndSepPunct{\mcitedefaultmidpunct}
{\mcitedefaultendpunct}{\mcitedefaultseppunct}\relax
\EndOfBibitem
\bibitem[CWJBeenakker(2013)]{CWJBeenakker1}
CWJBeenakker, \emph{Annu.\ Rev.\ Condens.\ Matter Phys.} \textbf{2013},
  \emph{4}, 113\relax
\mciteBstWouldAddEndPuncttrue
\mciteSetBstMidEndSepPunct{\mcitedefaultmidpunct}
{\mcitedefaultendpunct}{\mcitedefaultseppunct}\relax
\EndOfBibitem
\bibitem[Aasen \latin{et~al.}(2016)Aasen, Hell, Mishmash, Higginbotham, Danon,
  Leijnse, Jespersen, Folk, Marcus, Flensberg, and Alicea]{DAasen}
Aasen,~D.; Hell,~M.; Mishmash,~R.~V.; Higginbotham,~A.; Danon,~J.; Leijnse,~M.;
  Jespersen,~T.~S.; Folk,~J.~A.; Marcus,~C.~M.; Flensberg,~K.; Alicea,~J.
  \emph{Phys.\ Rev.\ X} \textbf{2016}, \emph{6}, 031016\relax
\mciteBstWouldAddEndPuncttrue
\mciteSetBstMidEndSepPunct{\mcitedefaultmidpunct}
{\mcitedefaultendpunct}{\mcitedefaultseppunct}\relax
\EndOfBibitem
\bibitem[Sau \latin{et~al.}(2010)Sau, Lutchyn, Tewari, and Sarma]{JDSau}
Sau,~J.~D.; Lutchyn,~R.~M.; Tewari,~S.; Sarma,~S.~D. \emph{Phys.\ Rev.\ Lett.}
  \textbf{2010}, \emph{104}, 040502\relax
\mciteBstWouldAddEndPuncttrue
\mciteSetBstMidEndSepPunct{\mcitedefaultmidpunct}
{\mcitedefaultendpunct}{\mcitedefaultseppunct}\relax
\EndOfBibitem
\bibitem[Mourik \latin{et~al.}(2012)Mourik, Zuo, Frolov, Plissard, Bakkers, and
  Kouwenhoven]{VMourik}
Mourik,~V.; Zuo,~K.; Frolov,~S.~M.; Plissard,~S.~R.; Bakkers,~E. P. A.~M.;
  Kouwenhoven,~L.~P. \emph{Science} \textbf{2012}, \emph{336}, 1003--1007\relax
\mciteBstWouldAddEndPuncttrue
\mciteSetBstMidEndSepPunct{\mcitedefaultmidpunct}
{\mcitedefaultendpunct}{\mcitedefaultseppunct}\relax
\EndOfBibitem
\bibitem[Deng \latin{et~al.}(2016)Deng, Vaitiek\.{e}nas, Hansen, Danon,
  Leijnse, Flensberg, Nyg\.{a}rd, Krogstrup, and Marcus]{MTDeng}
Deng,~M.~T.; Vaitiek\.{e}nas,~S.; Hansen,~E.~B.; Danon,~J.; Leijnse,~M.;
  Flensberg,~K.; Nyg\.{a}rd,~J.; Krogstrup,~P.; Marcus,~C.~M. \emph{Science}
  \textbf{2016}, \emph{354}, 1557--1562\relax
\mciteBstWouldAddEndPuncttrue
\mciteSetBstMidEndSepPunct{\mcitedefaultmidpunct}
{\mcitedefaultendpunct}{\mcitedefaultseppunct}\relax
\EndOfBibitem
\bibitem[Nadj-Perge \latin{et~al.}(2014)Nadj-Perge, Drozdov, J.~Li, Jeon, Seo,
  MacDonald, Bernevig, and Yazdani]{SNadjPerge}
Nadj-Perge,~S.; Drozdov,~I.~K.; J.~Li,~H.~C.; Jeon,~S.; Seo,~J.;
  MacDonald,~A.~H.; Bernevig,~B.~A.; Yazdani,~A. \emph{Science} \textbf{2014},
  \emph{346}, 602607\relax
\mciteBstWouldAddEndPuncttrue
\mciteSetBstMidEndSepPunct{\mcitedefaultmidpunct}
{\mcitedefaultendpunct}{\mcitedefaultseppunct}\relax
\EndOfBibitem
\bibitem[Ruby \latin{et~al.}(2017)Ruby, Heinrich, Peng, and von Oppen K.
  J.~Franke]{MRuby}
Ruby,~M.; Heinrich,~B.~W.; Peng,~Y.; von Oppen K. J.~Franke,~F. \emph{Nano.\
  Lett.} \textbf{2017}, \emph{17}, 4473--4477\relax
\mciteBstWouldAddEndPuncttrue
\mciteSetBstMidEndSepPunct{\mcitedefaultmidpunct}
{\mcitedefaultendpunct}{\mcitedefaultseppunct}\relax
\EndOfBibitem
\bibitem[Kim \latin{et~al.}(2018)Kim, Palacio-Morales, Posske, R\'{o}zsa,
  Palot\'{a}s, Szunyogh, Thorwart, and Wiesendanger]{HKim}
Kim,~H.; Palacio-Morales,~A.; Posske,~T.; R\'{o}zsa,~L.; Palot\'{a}s,~K.;
  Szunyogh,~L.; Thorwart,~M.; Wiesendanger,~R. \emph{Sci.\ Adv.} \textbf{2018},
  \emph{4}, eaar5251\relax
\mciteBstWouldAddEndPuncttrue
\mciteSetBstMidEndSepPunct{\mcitedefaultmidpunct}
{\mcitedefaultendpunct}{\mcitedefaultseppunct}\relax
\EndOfBibitem
\bibitem[Palacio-Morales \latin{et~al.}(2019)Palacio-Morales, Mascot, Cocklin,
  Kim, Rachel, Morr, and Wiesendanger]{APMorales}
Palacio-Morales,~A.; Mascot,~E.; Cocklin,~S.; Kim,~H.; Rachel,~S.; Morr,~D.~K.;
  Wiesendanger,~R. \emph{Sci.\ Adv.} \textbf{2019}, \emph{5}, eaav6600\relax
\mciteBstWouldAddEndPuncttrue
\mciteSetBstMidEndSepPunct{\mcitedefaultmidpunct}
{\mcitedefaultendpunct}{\mcitedefaultseppunct}\relax
\EndOfBibitem
\bibitem[M\'{e}nard \latin{et~al.}(2019)M\'{e}nard, Mesaros, Brun,
  Debontridder, Roditchev, Simon, and Cren]{GCMenard}
M\'{e}nard,~G.~C.; Mesaros,~A.; Brun,~C.; Debontridder,~F.; Roditchev,~D.;
  Simon,~P.; Cren,~T. \emph{Nat.\ Commun.} \textbf{2019}, \emph{10}, 2587\relax
\mciteBstWouldAddEndPuncttrue
\mciteSetBstMidEndSepPunct{\mcitedefaultmidpunct}
{\mcitedefaultendpunct}{\mcitedefaultseppunct}\relax
\EndOfBibitem
\bibitem[Schmid \latin{et~al.}(1996)Schmid, Hebenstreit, Varga, and
  Crampin]{MSchmid}
Schmid,~M.; Hebenstreit,~W.; Varga,~P.; Crampin,~S. \emph{Phys.\ Rev.\ Lett.}
  \textbf{1996}, \emph{76}, 2298--2301\relax
\mciteBstWouldAddEndPuncttrue
\mciteSetBstMidEndSepPunct{\mcitedefaultmidpunct}
{\mcitedefaultendpunct}{\mcitedefaultseppunct}\relax
\EndOfBibitem
\bibitem[M\"{u}ller \latin{et~al.}(2016)M\"{u}ller, N\'{e}el, Crampin, and
  Kr\"{o}ger]{MMuller}
M\"{u}ller,~M.; N\'{e}el,~N.; Crampin,~S.; Kr\"{o}ger,~J. \emph{Phys.\ Rev.\
  Lett.} \textbf{2016}, \emph{117}, 136803\relax
\mciteBstWouldAddEndPuncttrue
\mciteSetBstMidEndSepPunct{\mcitedefaultmidpunct}
{\mcitedefaultendpunct}{\mcitedefaultseppunct}\relax
\EndOfBibitem
\bibitem[Harrison \latin{et~al.}(2004)Harrison, Woodruff, and
  Robinson]{MJHarrison}
Harrison,~M.~J.; Woodruff,~D.~P.; Robinson,~J. \emph{Sur.\ Sci.} \textbf{2004},
  \emph{572}, 309--317\relax
\mciteBstWouldAddEndPuncttrue
\mciteSetBstMidEndSepPunct{\mcitedefaultmidpunct}
{\mcitedefaultendpunct}{\mcitedefaultseppunct}\relax
\EndOfBibitem
\bibitem[Li \latin{et~al.}(2007)Li, Xiao, Zu, and Dong]{DFLi}
Li,~D.~F.; Xiao,~H.~Y.; Zu,~X.~T.; Dong,~H.~N. \emph{Physica B} \textbf{2007},
  \emph{392}, 217--220\relax
\mciteBstWouldAddEndPuncttrue
\mciteSetBstMidEndSepPunct{\mcitedefaultmidpunct}
{\mcitedefaultendpunct}{\mcitedefaultseppunct}\relax
\EndOfBibitem
\bibitem[Libra \latin{et~al.}(2007)Libra, Veltrusk\'{a}, and
  Matol\'{i}n]{JLibra}
Libra,~J.; Veltrusk\'{a},~K.; Matol\'{i}n,~V. \emph{Phys.\ Rev.\ B}
  \textbf{2007}, \emph{76}, 165438\relax
\mciteBstWouldAddEndPuncttrue
\mciteSetBstMidEndSepPunct{\mcitedefaultmidpunct}
{\mcitedefaultendpunct}{\mcitedefaultseppunct}\relax
\EndOfBibitem
\bibitem[Zhang and Yang(2018)Zhang, and Yang]{YXZhang}
Zhang,~Y.; Yang,~Z. \emph{Sur.\ Sci.} \textbf{2018}, \emph{670}, 68--71\relax
\mciteBstWouldAddEndPuncttrue
\mciteSetBstMidEndSepPunct{\mcitedefaultmidpunct}
{\mcitedefaultendpunct}{\mcitedefaultseppunct}\relax
\EndOfBibitem
\bibitem[Yu \latin{et~al.}(2017)Yu, Fu, She, Lu, Guo, Li, Meng, and Cao]{YYu}
Yu,~Y.; Fu,~H.; She,~L.; Lu,~S.; Guo,~Q.; Li,~H.; Meng,~S.; Cao,~G. \emph{ACS
  Nano} \textbf{2017}, \emph{11}, 2143--2149\relax
\mciteBstWouldAddEndPuncttrue
\mciteSetBstMidEndSepPunct{\mcitedefaultmidpunct}
{\mcitedefaultendpunct}{\mcitedefaultseppunct}\relax
\EndOfBibitem
\bibitem[Kim \latin{et~al.}(2011)Kim, Chua, Fiete, Nam, MacDonald, and
  Shih]{JKim}
Kim,~J.; Chua,~V.; Fiete,~G.~A.; Nam,~H.; MacDonald,~A.~H.; Shih,~C.~K.
  \emph{Phys.\ Rev.\ B} \textbf{2011}, \emph{84}, 014517\relax
\mciteBstWouldAddEndPuncttrue
\mciteSetBstMidEndSepPunct{\mcitedefaultmidpunct}
{\mcitedefaultendpunct}{\mcitedefaultseppunct}\relax
\EndOfBibitem
\bibitem[Kim \latin{et~al.}(2012)Kim, Chua, Fiete, Nam, MacDonald, and
  Shih]{JKim1}
Kim,~J.; Chua,~V.; Fiete,~G.~A.; Nam,~H.; MacDonald,~A.~H.; Shih,~C.~K.
  \emph{Nat.\ Phys.} \textbf{2012}, \emph{8}, 464\relax
\mciteBstWouldAddEndPuncttrue
\mciteSetBstMidEndSepPunct{\mcitedefaultmidpunct}
{\mcitedefaultendpunct}{\mcitedefaultseppunct}\relax
\EndOfBibitem
\bibitem[Kim \latin{et~al.}(2002)Kim, Thalmeier, and Won]{KMaki}
Kim,~K.; Thalmeier,~P.; Won,~H. \emph{Phys.\ Rev.\ B} \textbf{2002}, \emph{65},
  140502(R)\relax
\mciteBstWouldAddEndPuncttrue
\mciteSetBstMidEndSepPunct{\mcitedefaultmidpunct}
{\mcitedefaultendpunct}{\mcitedefaultseppunct}\relax
\EndOfBibitem
\bibitem[Watanabe \latin{et~al.}(2004)Watanabe, Nohara, Hanaguri, and
  Takagi]{TWatanabe}
Watanabe,~T.; Nohara,~M.; Hanaguri,~T.; Takagi,~H. \emph{Phys.\ Rev.\ Lett.}
  \textbf{2004}, \emph{92}, 147002)\relax
\mciteBstWouldAddEndPuncttrue
\mciteSetBstMidEndSepPunct{\mcitedefaultmidpunct}
{\mcitedefaultendpunct}{\mcitedefaultseppunct}\relax
\EndOfBibitem
\bibitem[Jiao \latin{et~al.}(2014)Jiao, Tang, Sun, Liu, Tao, Feng, Zeng, Xu,
  and Cao]{WHJiao}
Jiao,~W.~H.; Tang,~Z.~T.; Sun,~Y.~L.; Liu,~Y.; Tao,~Q.; Feng,~C.~M.;
  Zeng,~Y.~W.; Xu,~Z.~A.; Cao,~G.~H. \emph{J.\ Am.\ Chem.\ Soc.} \textbf{2014},
  \emph{136}, 1284--1287)\relax
\mciteBstWouldAddEndPuncttrue
\mciteSetBstMidEndSepPunct{\mcitedefaultmidpunct}
{\mcitedefaultendpunct}{\mcitedefaultseppunct}\relax
\EndOfBibitem
\bibitem[Du \latin{et~al.}(2015)Du, Fang, Wang, Li, Du, Yang, Zhu, and
  Wen]{ZDu}
Du,~Z.; Fang,~D.; Wang,~Z.; Li,~Y.; Du,~G.; Yang,~H.; Zhu,~X.; Wen,~H.~H.
  \emph{Sci.\ Rep.} \textbf{2015}, \emph{5}, 9408)\relax
\mciteBstWouldAddEndPuncttrue
\mciteSetBstMidEndSepPunct{\mcitedefaultmidpunct}
{\mcitedefaultendpunct}{\mcitedefaultseppunct}\relax
\EndOfBibitem
\bibitem[Prozorov \latin{et~al.}(2006)Prozorov, Olheiser, Giannetta, Uozato,
  and Tamegai]{RProzorov}
Prozorov,~R.; Olheiser,~T.~A.; Giannetta,~R.~W.; Uozato,~K.; Tamegai,~T.
  \emph{Phys.\ Rev.\ B} \textbf{2006}, \emph{73}, 184523\relax
\mciteBstWouldAddEndPuncttrue
\mciteSetBstMidEndSepPunct{\mcitedefaultmidpunct}
{\mcitedefaultendpunct}{\mcitedefaultseppunct}\relax
\EndOfBibitem
\bibitem[Yasuzuka \latin{et~al.}(2020)Yasuzuka, Uji, Sugiura, Terashima,
  Nogami, Ichimura, and Tanda]{SYasuzuka}
Yasuzuka,~S.; Uji,~S.; Sugiura,~S.; Terashima,~T.; Nogami,~Y.; Ichimura,~K.;
  Tanda,~S. \emph{J.\ Supercond.\ Nov.\ Magn.} \textbf{2020}, \emph{33},
  953--958\relax
\mciteBstWouldAddEndPuncttrue
\mciteSetBstMidEndSepPunct{\mcitedefaultmidpunct}
{\mcitedefaultendpunct}{\mcitedefaultseppunct}\relax
\EndOfBibitem
\bibitem[Trainer \latin{et~al.}(2020)Trainer, Wang, Bobba, Samuelson, Xi,
  Zasadzinski, Nieminen, Bansil, and Iavarone]{DJTrainer}
Trainer,~D.~J.; Wang,~B.; Bobba,~F.; Samuelson,~N.; Xi,~X.; Zasadzinski,~J.;
  Nieminen,~J.; Bansil,~A.; Iavarone,~M. \emph{ACS Nano.} \textbf{2020},
  \emph{14}, 2718--2728\relax
\mciteBstWouldAddEndPuncttrue
\mciteSetBstMidEndSepPunct{\mcitedefaultmidpunct}
{\mcitedefaultendpunct}{\mcitedefaultseppunct}\relax
\EndOfBibitem
\bibitem[Bardeen \latin{et~al.}(1957)Bardeen, Cooper, and Schrieffer]{JBardeen}
Bardeen,~J.; Cooper,~L.~N.; Schrieffer,~J.~R. \emph{Phys.\ Rev.\ B}
  \textbf{1957}, \emph{108}, 1175\relax
\mciteBstWouldAddEndPuncttrue
\mciteSetBstMidEndSepPunct{\mcitedefaultmidpunct}
{\mcitedefaultendpunct}{\mcitedefaultseppunct}\relax
\EndOfBibitem
\bibitem[Prozorov and Giannetta(2006)Prozorov, and Giannetta]{RProzorov1}
Prozorov,~R.; Giannetta,~R.~W. \emph{Supercond.\ Sci.\ Technol.} \textbf{2006},
  \emph{19}, R41\relax
\mciteBstWouldAddEndPuncttrue
\mciteSetBstMidEndSepPunct{\mcitedefaultmidpunct}
{\mcitedefaultendpunct}{\mcitedefaultseppunct}\relax
\EndOfBibitem
\bibitem[Tinkham(1996)]{MTinkham}
Tinkham,~M. \emph{Introduction to Superconductivity (Dover Publications,
  Mineola, New York.} \textbf{1996}, 62--63\relax
\mciteBstWouldAddEndPuncttrue
\mciteSetBstMidEndSepPunct{\mcitedefaultmidpunct}
{\mcitedefaultendpunct}{\mcitedefaultseppunct}\relax
\EndOfBibitem
\bibitem[Bose \latin{et~al.}(2010)Bose, Garcia-Garcia, Ugeda, Urbina,
  Michaelis, Brihuega, and Kern]{SBose}
Bose,~S.; Garcia-Garcia,~A.~M.; Ugeda,~M.~M.; Urbina,~J.~D.; Michaelis,~C.~H.;
  Brihuega,~I.; Kern,~K. \emph{Nat.\ Mater.} \textbf{2010}, \emph{9}, 550\relax
\mciteBstWouldAddEndPuncttrue
\mciteSetBstMidEndSepPunct{\mcitedefaultmidpunct}
{\mcitedefaultendpunct}{\mcitedefaultseppunct}\relax
\EndOfBibitem
\bibitem[Zhang \latin{et~al.}(2010)Zhang, Cheng, Li, Sun, Wang, Zhu, He, Wang,
  Ma, Chen, Wang, Liu, Lin, Jia, and Xue]{TZhang}
Zhang,~T.; Cheng,~P.; Li,~W.~J.; Sun,~Y.~J.; Wang,~G.; Zhu,~X.~G.; He,~K.;
  Wang,~L.; Ma,~X.; Chen,~X.; Wang,~Y.; Liu,~Y.; Lin,~H.~Q.; Jia,~J.~F.;
  Xue,~Q.~K. \emph{Nat.\ Phys.} \textbf{2010}, \emph{6}, 104--108\relax
\mciteBstWouldAddEndPuncttrue
\mciteSetBstMidEndSepPunct{\mcitedefaultmidpunct}
{\mcitedefaultendpunct}{\mcitedefaultseppunct}\relax
\EndOfBibitem
\bibitem[Cherkez \latin{et~al.}(2014)Cherkez, Cuevas, Brun, Cren, M\'{e}nard,
  Debontridder, Stolyarov, and Roditche]{VCherkez}
Cherkez,~V.; Cuevas,~J.~C.; Brun,~C.; Cren,~T.; M\'{e}nard,~G.;
  Debontridder,~F.; Stolyarov,~V.~S.; Roditche,~D. \emph{Phys.\ Rev.\ X}
  \textbf{2014}, \emph{4}, 011033\relax
\mciteBstWouldAddEndPuncttrue
\mciteSetBstMidEndSepPunct{\mcitedefaultmidpunct}
{\mcitedefaultendpunct}{\mcitedefaultseppunct}\relax
\EndOfBibitem
\bibitem[Lindemann(1910)]{FLindemann}
Lindemann,~F. \emph{Z.\ Phys.} \textbf{1910}, \emph{11}, 609\relax
\mciteBstWouldAddEndPuncttrue
\mciteSetBstMidEndSepPunct{\mcitedefaultmidpunct}
{\mcitedefaultendpunct}{\mcitedefaultseppunct}\relax
\EndOfBibitem
\bibitem[Khrapak(2020)]{SAKhrapak}
Khrapak,~S.~A. \emph{Phys.\ Rev.\ Research} \textbf{2020}, \emph{2},
  012040(R)\relax
\mciteBstWouldAddEndPuncttrue
\mciteSetBstMidEndSepPunct{\mcitedefaultmidpunct}
{\mcitedefaultendpunct}{\mcitedefaultseppunct}\relax
\EndOfBibitem
\bibitem[Si \latin{et~al.}(2013)Si, Liu, Duan, and Liu]{CSi}
Si,~C.; Liu,~Z.; Duan,~W.; Liu,~F. \emph{Phys.\ Rev.\ Lett.} \textbf{2013},
  \emph{111}, 196802\relax
\mciteBstWouldAddEndPuncttrue
\mciteSetBstMidEndSepPunct{\mcitedefaultmidpunct}
{\mcitedefaultendpunct}{\mcitedefaultseppunct}\relax
\EndOfBibitem
\bibitem[Ge \latin{et~al.}(2015)Ge, Wan, Yang, and Yao]{YGe}
Ge,~Y.; Wan,~W.; Yang,~F.; Yao,~Y. \emph{New J.\ Phys.} \textbf{2015},
  \emph{17}, 035008\relax
\mciteBstWouldAddEndPuncttrue
\mciteSetBstMidEndSepPunct{\mcitedefaultmidpunct}
{\mcitedefaultendpunct}{\mcitedefaultseppunct}\relax
\EndOfBibitem
\bibitem[Li \latin{et~al.}(2018)Li, Zhao, Zeng, Zulfiqar, and Ni]{GLi}
Li,~G.; Zhao,~Y.; Zeng,~S.; Zulfiqar,~M.; Ni,~J. \emph{J.\ Phys.\ Chem.\ C}
  \textbf{2018}, \emph{122}, 16916--16924\relax
\mciteBstWouldAddEndPuncttrue
\mciteSetBstMidEndSepPunct{\mcitedefaultmidpunct}
{\mcitedefaultendpunct}{\mcitedefaultseppunct}\relax
\EndOfBibitem
\bibitem[Woodruff and Robinson(2003)Woodruff, and Robinson]{DPWoodruff}
Woodruff,~D.~P.; Robinson,~J. \emph{Appl.\ Sur.\ Sci.} \textbf{2003},
  \emph{219}, 1--10\relax
\mciteBstWouldAddEndPuncttrue
\mciteSetBstMidEndSepPunct{\mcitedefaultmidpunct}
{\mcitedefaultendpunct}{\mcitedefaultseppunct}\relax
\EndOfBibitem
\end{mcitethebibliography}

\end{document}